\documentclass[onecolumn,preprintnumbers,amsmath,amssymb]{revtex4-1}
\usepackage{graphicx}
\usepackage{dcolumn}
\usepackage{bm}

\begin{document}

\title{Cooperative dynamics in the Fiber Bundle Model}
\author{Bikas K. Chakrabarti\email{bikask.chakrabarti@saha.ac.in}}
\affiliation{Saha Institute of Nuclear Physics, Kolkata 700064, India} 
\affiliation{S. N. 
Bose National Centre for Basic Sciences, Kolkata 700106, India.} 
\author{Soumyajyoti Biswas\email{soumyajyotibiswas@gmail.com}} 
\affiliation{SRM University-AP, Andhra Pradesh - 522502, India.}
\author{Srutarshi Pradhan\email{srutarshi.pradhan@ntnu.no}} 
\affiliation{PoreLab, Department of Physics, Norwegian University of
Science and Technology, NO--7491 Trondheim, Norway.}

\begin{abstract}
 We discuss the cooperative failure dynamics in the
Fiber Bundle Model where the individual  elements or fibers
are Hookean springs, having identical spring constant  but different
breaking strengths. When the bundle is stressed or strained,
especially in the equal-load-sharing scheme, the load
supported by the failed  fiber 
gets shared equally by the rest of the surviving fibers. This
mean-field type statistical feature  (absence of fluctuations)
in the load-sharing mechanism helped major analytical 
developments in the study
of breaking dynamics in the model and precise comparisons
with simulation results. We intend to present  a brief review
on these developments.
\end{abstract}
\maketitle
\section{Introduction}
Fiber bundle model (FBM) has been used widely for studying fracture 
and failure \cite{RMP} 
of composite materials under external loading. The simplicity  
of the model allow us to achieve analytic 
solutions \cite{pc01,pbc02,bpc03}
 to an  extent that is not possible in any other fracture models. 
For these very reasons, FBM is widely used as a model of breakdown that extends beyond 
disordered solids. In fact,
FBM was first introduced in connection with textile 
engineering \cite{p26}. Physicists took interest in it recently 
to explore the 
critical failure dynamics and avalanche phenomena during such stress-induced 
failures \cite{HH,PHH,ph07,dd07}. Apart from
the classical fracture-failure in composites, FBM has been 
used successfully for studying noise-induced (creep/fatigue) failure 
\cite{Lawn,Coleman,Ciliberto,Roux,pc03} where a fixed
load is applied on the system and external noise triggers the 
failure of elements. 
Furthermore, it was used as a model for other geophysical phenomena, such as 
snow avalanche \cite{reiw_2009}, land slides \cite{cohen_2009, pollen_2005}, biological materials \cite{pugno} or even earthquakes \cite{sorr_92}.
In this review article we want to concentrate only on 
the cooperative dynamical aspects in FBM. 

\begin{figure}[t]
\begin{center}
\includegraphics[width=6cm]{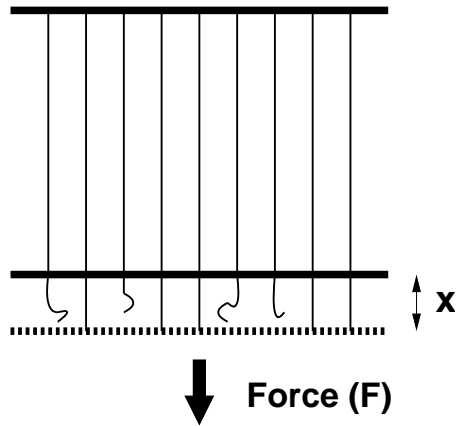}
\vspace{0.6cm}
\caption {\label{fig:FBM-model}
A cartoon of the Fiber Bundle Model where a macroscopically
large number ($N$) of Hookean springs, with identical spring
constant but different breaking thresholds hang parallelly
from an upper rigid bar and a load/force $F$ is applied at the lower horizontal
rigid bar (not allowing any local deformation of the bar
and consequent local stress concentration). If any spring
fails at any time, the (extra) load is shared by the
surviving fibers at that time. In the equal-load-sharing
scheme, considered here, this extra load is shared equally
by all the surviving fibers ($x$ denoting the strain of
the surviving fibers).}
 
\end{center}
\end{figure}
F. T. Peirce, a textile engineer, introduced the Fiber Bundle Model \cite{p26} 
in $1926$ to study the strength of cotton yarns.
Later, in $1945$ Daniels discussed some static behavior of such a bundle 
\cite{d45} and the model was brought to the attention of 
physicists  in $1989$ by Sornette \cite{s89} who started analysing the failure 
process. 
Even though FBM was designed initially as a model for
fracture or failure
of a set of parallel elements (fibers), having different breaking 
thresholds, with the collective load-sharing scheme, 
 the failure dynamics in the model shows
all the attributes of the critical phenomena and associated phase
transition. It seems, due to the usefulness and richness, FBM plays the same 
role (in the field of fracture) as the Ising model in
magnetism \cite{i25}. 

In FBM, a  number of parallel Hookean
springs or fibers are clamped between two horizontal platforms 
(Figure \ref{fig:FBM-model}). The breaking strengths of the springs or fibers 
are different. 
When the load per fiber (stress) exceeds a fiber's  
own threshold, it fails. 
The load it carried has to be shared by the
surviving fibers. If the lower platform deforms under loading while upper 
platform is rigid, fibers in the neighborhood of just-failed fiber will 
absorb more of the load compared to 
fibers sitting further away and this arrangement is called local-load-sharing 
(LLS) scheme \cite{hp91,gip93}. If both the platforms are rigid, the load has 
to equally distributed among 
all the surviving fibers, which is called the equal-load-sharing (ELS) scheme. 
Intermediate load redistribution schemes are also studied 
(see e.g., \cite{hkh02}) where a part of the load is shared locally within 
a few fibers and the 
rest is shared globally among all the fibers.

How does cooperative dynamics set in? 
In case of ELS, all the intact fibers carry the load equally. When a fiber 
fails, the stress level increases on the remaining fibers and that can trigger 
more fiber-failures (successive failure). As long as the initial
load is low, the successive failures of the fibers 
remain small and though the
strain  (stretch) of the bundle grow with increasing stress (load), the bundle 
as a
whole does not fail. Once the initial load reaches  a ``critical" value, 
determined by the fiber strength distribution, the successive
failures become a global (catastrophic) one and the bundle collapses. 

We arrange this review article as follows: In the short introduction
(section I), we elaborate the concept of the fiber bundle model and its 
evolution as a
fracture model. Section II deals with the equal-load-sharing FBM where 
 we demonstrate the dynamical behavior in FBM with
evolution dynamics and their solutions. 
Analytic results are compared with numerical simulations in this
 section. In section III we discuss noise-induced failure dynamics in 
 FBM through theory, simulation and real-data analysis. The self-organizing 
 mechanism in FBM is discussed in section IV. We reserve section V for 
 discussions on some works which would help
to understand the cooperative dynamics in FBM. Finally, we keep a short Summary and Conclusion section (section VI) at the end.

\section{Equal Load Sharing FBM}

We consider a FBM having 
$N$ parallel fibers placed between two rigid bars.  Each fiber follows
Hook's law with a force $f$ to the stretch value $x$ as $f=\kappa x\;$,
where $\kappa$ is the spring constant.  To make things simpler, we consider 
$\kappa=1$ for all the fibers. 
Each fiber has a particular strength threshold value and if the 
stretch $x$ exceeds this threshold, the fiber fails irreversibly.  We are 
interested in the 
equal-load-sharing (ELS) mode (the bars are 
rigid), and by construction of the model, the applied load has to be shared 
equally by the intact fibers. 

Other than the analytical treatment of the model, several aspects of the model 
are also explored numerically.
The implementation of the model, particularly in the equal load sharing version we discuss here, is straightforward. The load is initially applied
to each fiber equally. The fibers having failure thresholds less than the applied load are irreversibly broken. The load carried by those fibers is
redistributed equally among the remaining fibers, which can cause further 
breaking. The redistribution continue until no new fibers are breaking. 
The external load is held constant during the whole redistribution process. 
This is due to the separation of time scales of externally applied loading 
rate and the
internal (elastic) relaxation processes within materials. After the end of 
each redistribution cycle, the external load is further increased to continue 
the dynamics. 
This process continues until the entire system is broken. The critical 
strength, avalanche statistics and other critical exponents are calculated 
from this dynamics,
which as will see, match well with the analytical results.

\subsection{Fiber strength distributions}
\begin{figure}[h]
\begin{center}
\includegraphics[width=8cm]{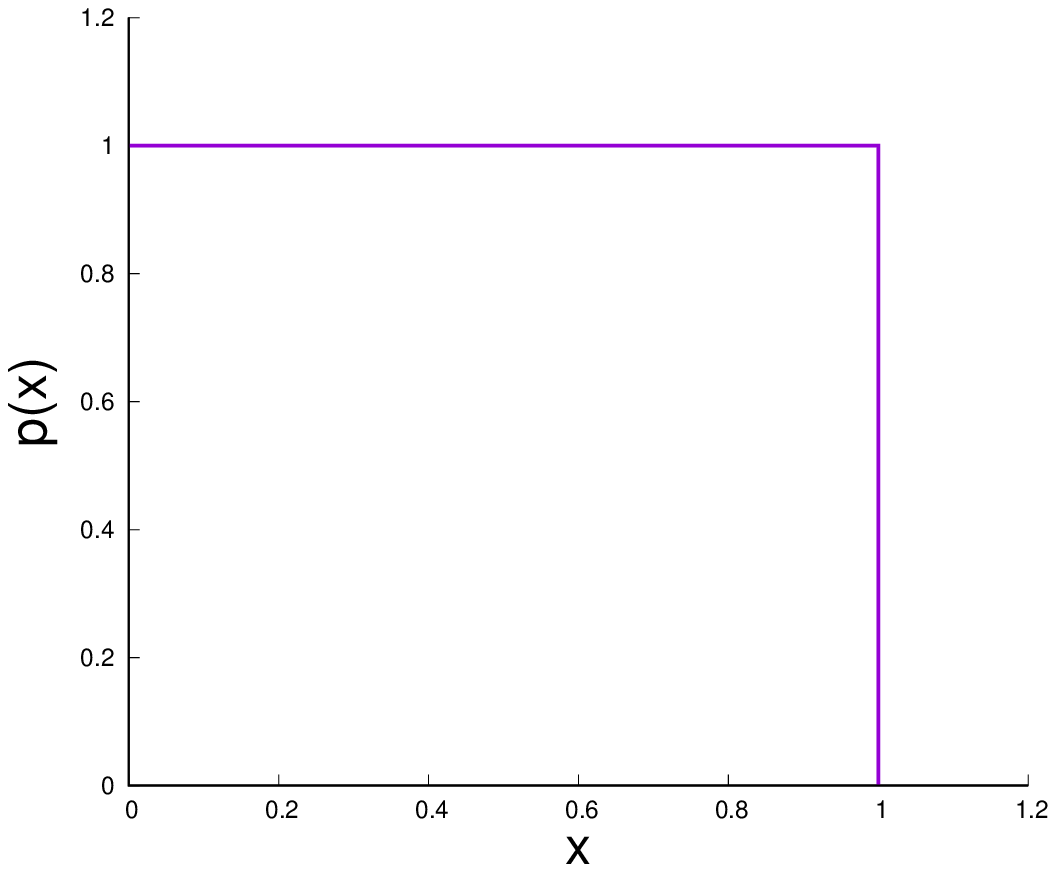}
\includegraphics[width=8cm]{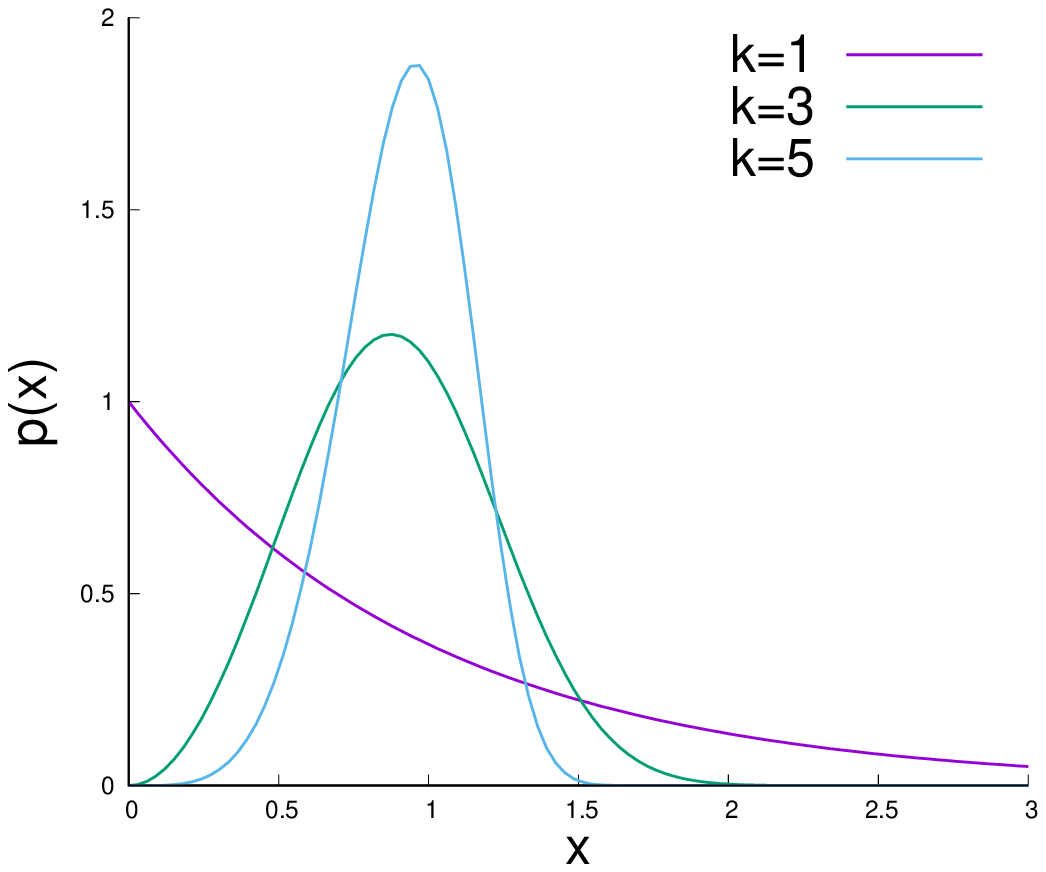}
\caption {\label{fig:Weibull-dists}
The uniform and Weibull distributions of fiber strengths (thresholds).}
\end{center}
\end{figure}

The fiber strength thresholds
are drawn from a probability density $p(x)$. The corresponding cumulative probability is
\begin{equation}
\label{eq2}
P(x)=\int_0^x\  p(y) dy\;.
\end{equation}
The most used threshold distributions are  uniform and 
Weibull distributions (see Figure \ref{fig:Weibull-dists}) in FBM literature. 

For a uniform distribution we can write 
\begin{equation}
\label{eq:uniform}
p(x)=1;   P(x)=x,
\end{equation}
where the range of function is between $0$ to $1$.
The cumulative
Weibull distribution has a form:
\begin{equation}
\label{eq:Weibull-cumm}
P(x)= 1-\exp(-x^k),
\end{equation}
where, $k$ is the shape parameter or Weibull index. 
The corresponding probability distribution takes the form:
\begin{equation}
\label{eq:Weibull-prob}
p(x)=kx^{k-1}\exp(-x^k).
\end{equation}
The shape of the uniform and Weibull distributions are shown in Figure (\ref{fig:Weibull-dists}). 
The range of definition is between 0 to $\infty$.

\subsection{The critical values}

When we stretch the bundle by applying a force, the fibers fail according to 
their thresholds, 
the weakest first, then the next weakest and so on. 
If  $N_f$ fibers have failed at a stretch value 
$x$, the force on the bundle is
\begin{equation}
\label{eq:load-curve}
F=(N-N_f)x=N(1-P(x))x,
\end{equation}
as $\kappa=1$.  
The normalized force ($F/N$) vs. stretch $x$ curve
looks like a parabola (Figure \ref{fig:phase-uniform-Weibull}).


It is obvious that the maximum of the force value is the strength of the 
bundle and the corresponding stretch 
value  ($x_c$) is called the critical stretch beyond which the bundle 
collapses.  
Therefore, we can define two distinct phases of the system: stable phase for 
$0<x\le x_c$ and unstable phase for $x > x_c$.

The critical stretch value can be obtained easily by setting $dF(x)/dx =0$:
\begin{equation}
\label{eq13}
1-x_c p(x_c)-P(x_c)=0.
\end{equation}
\subsubsection{Uniform threshold distribution}
 Substituting the $p(x_c)$ and $P(x_c)$ values for  
 uniform distribution, we obtain  
\begin{equation}
\label{eq-delta_c}
x_c =\left(\frac{1}{2}\right).
\end{equation}
Now inserting the $x_c$ value in the force expression 
(Eq. \ref{eq:load-curve}), we get 
\begin{equation}
\label{eq14}
\frac{F_c}{N}=\frac{1}{4}; 
\end{equation}
which is the critical strength of the bundle 
(Figure \ref{fig:phase-uniform-Weibull}).  

\subsubsection{Weibull threshold distribution}
\begin{figure}[t]
\begin{center}
\includegraphics[width=8cm]{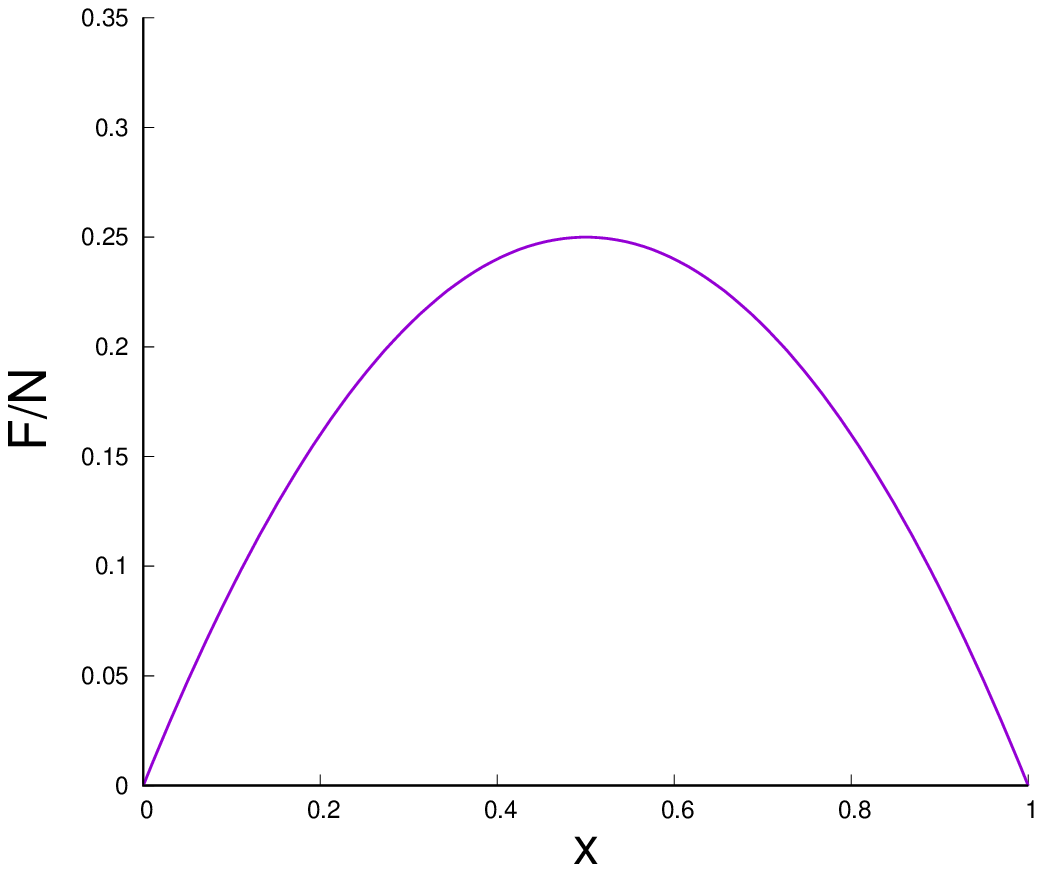}
\includegraphics[width=8cm]{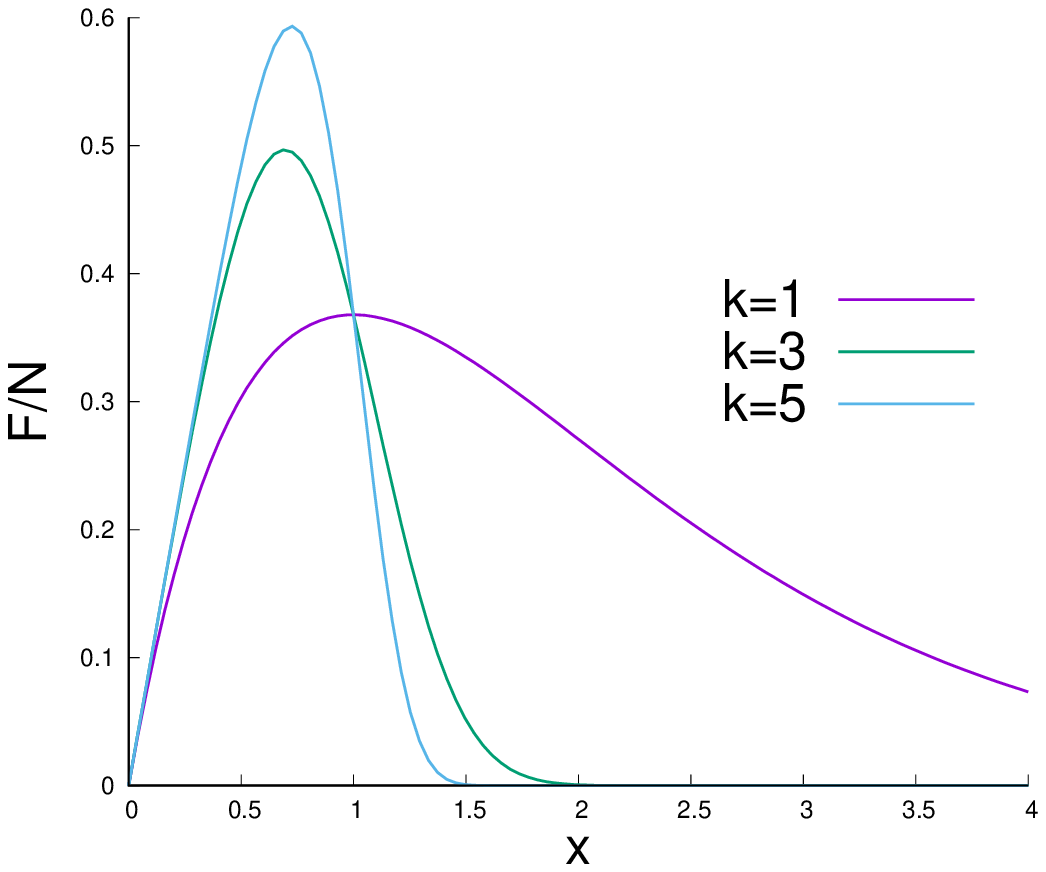}
\caption {\label{fig:phase-uniform-Weibull}
Normalized force ($F/N$) against extension $x$ for a fiber 
bundle with uniform ($x_c  = 0.5$) and Weibull ($x_c = 1, (1/3)^{1/3}, (1/5)^{1/5}$ for $k=1,3,5$ respectively) distributions of  strengths (thresholds).}
\end{center}
\end{figure}

In case of  Weibull distribution, 
at the force-maximum, insering the $P(x)$, $p(x)$ values into  
expression (Eq. \ref{eq13}), 
we obtain
\begin{equation}
\label{eq-Weibull-Delta_c1}
\exp(-x_c^k) - (x_c kx_c^{k-1}\exp(-x_c^k)) = 0.
\end{equation}
One can get the critical stretch value: 
\begin{equation}
\label{eq-Weibull-Delta_c2}
x_c =k^{-\frac{1}{k}};
\end{equation}
and the corresponding critical force value
\begin{equation}
\label{eq-F_c-Weibull}
\frac{F_c}{N} = k^{-\frac{1}{k}}\exp({-\frac{1}{k}}).
\end{equation}
For $k=1$, $x_c=1.0$ and $\frac{F_c}{N}=\frac{1}{e}$ (Figure \ref{fig:phase-uniform-Weibull}).

\subsection{Different ways of loading}

Now we will discuss how  the load or stress on the bundle can be applied.
In the FBM literature, the most common loading mechanism 
discussed \cite{hhp15,pc19} is the “weakest-link-failure” mechanism of loading.
This loading process ensures a separation in time scales between external loading and
internal stress redistribution. This is equivalent to a “quasi-static”
approach and noise/fluctuation in the threshold distribution
  influences the breaking dynamics as well as the avalanche statistics.

A fiber bundle can also be loaded in a different
way by applying a fixed amount of load at a time. In that case, all
fibers having failure threshold below the applied load, fail.
The stress on the surviving fibers then increases due to load redistribution.
The increased stress may drives further failures, and so on. This iterative
breaking process continues until an equilibrium is reached
where the intact fibers (those who can support the load)
is reached. 
One can also study the failure dynamics of the
bundle when the external load on the bundle is then increased
infinitesimally, but by a fixed amount (irrespective of the
fluctuations in the fiber strength distribution as discussed
above). 
Indeed, as shown recently
in Biswas and Chakrabarti \cite{bc20}, the universality class
of the dynamics of such fixed loading (even for the same ELS
mode of load redistribution after individual  fiber failure)
will be different from that for the quasi-static (or weakest
link failure type) loading discussed above and is given by
the Flory statistics \cite{flory} for linear polymers, accommodating the
Kolmogorov type dispersion in turbulence \cite{kolm}.

\subsection{The cooperative dynamics}
We are going to discuss now the cooperative dynamical behavior of the
breaking processes for the  bundle loaded by fixed amount per step 
(following the formulations
in the References \cite{pc01,bpc03,ph07,RMP,hhp15,pc19,brc15}). 

Let us assume that an external force $F$ is applied to the fiber bundle.  
The stress on the bundle (the external load per fiber) is

\begin{equation}
\sigma = F/N.
\label{eq:2.4.0.1}\end{equation}
Let us call $N_t$ be the number of
surviving fibers after $t$ steps in the stress redistribution cycle, with $N_0 = N$. 

The effective stress  becomes
\begin{equation}
\sigma_t = N\sigma /N_t.
\label{eq:2.4.0.2}\end{equation}
Therefore, $N P(N\sigma/N_t)$ of fibers will fail in the first stress 
redistribution cycle. The number of intact fibers in the next cycle will be
\begin{equation}
 N_{t+1} = N - NP(N\sigma/N_t).
\label{eq:2.4.0.4}\end{equation}
Using $n_t = N_t/N$, Eq. (\ref{eq:2.4.0.4}) takes the form of a  recursion 
relation,   
\begin{equation}
n_{t+1} = 1 - P(\sigma/n_t), 
\label{eq:2.4.0.6}\end{equation}
with $\sigma$ as the control parameter and $n_0=1$ is the start value.

The character of an iterative dynamics is determined by 
its {\it fixed points} (denoted by *) where a dynamical variable remains 
exactly at the same value it had in the previous step of the dynamics. 
In other words, a fixed point is a value (of a dynamical variable) that is 
mapped onto itself by the iteration. 
The dynamics stops or it becomes locked at the fixed point. 

One can find out the
possible fixed points 
$n^*$  of (\ref{eq:2.4.0.6}), which satisfy
\begin{equation}
n^* = 1- P(\sigma/n^*),
\label{eq:2.4.1.1}\end{equation} 
 and the solutions of the breaking dynamics at the fixed point. 
\subsection{The critical exponents}
If we consider that the fiber strengths follow uniform  distribution, 
the recursion relation can be written as  
\begin{equation}
n_{t+1}= 1-\sigma/n_t.
\label{eq:recur-uni}
\end{equation}
Consequently, at the fixed point the relation assumes 
a simple form 
\begin{equation} 
(n^*)^2 -n^* +\sigma =0 ,
\label{eq:2.4.1.9}\end{equation}
with solution
\begin{equation}
n^* = {\textstyle \frac{1}{2}} \pm \left(\sigma_c-\sigma\right)^{1/2}.
\label{eq:2.4.1.10}\end{equation}
Here critical stress value $\sigma_c=1/4$, beyond which the bundle collapses 
completely.  
In (\ref{eq:2.4.1.10}) the upper sign gives $n^*>n_c$ which corresponds to 
a stable fixed point. 
From this solution, it is easy to derive the order parameter, susceptibility 
and relaxation time (all defined below). 

The fixed-point solution gives the critical value: 
($\sigma =\sigma_c$)
\begin{equation}
n_c^* = \frac{1}{2}.
\label{en:nc-uniform}\end{equation}
Therefore, the fixed-point solution can be presented as
\begin{equation}
n^*(\sigma)-n_c^* \propto (\sigma_c-\sigma)^{\beta}, \hspace{8mm} \beta = {\textstyle \frac{1}{2}}.
\label{eq:order-uni}\end{equation}
Clearly, $n^*(\sigma)-n_c^*$ can be considered like an {\it order parameter}i,
which shows a
clear transition from non-zero to zero value at $\sigma_c$.

The  {\it susceptibility} is defined as $\chi=-dn^*/d\sigma$ and the 
fixed-point solution gives
\begin{equation}
\chi \propto (\sigma_c-\sigma)^{-\gamma}, \hspace{8mm} \gamma = {\textstyle \frac{1}{2}};
\label{eq:kappa-uni}\end{equation}
 which follows a power law and diverges at the critical point $\sigma_c$.

The dynamical approach very near a fixed point is very interesting and  this 
can be investigated by expanding the differences $n_t-n^*$ around the fixed 
point. In case of uniform distribution, the recursion relation 
(\ref{eq:recur-uni}),  gives
\begin{equation}
n_{t+1}-n^* = \frac{\sigma}{n^*}-\frac{\sigma}{n_t}=\frac{\sigma}{n_tn^*}\;(n_t-n^*) \simeq \frac{\sigma}{n^{*2}} (n_t-n^*).
\label{eq:2.4.1.13}
\end{equation}
Clearly, the fixed point is approached with exponentially decreasing steps:
\begin{equation}
n_t-n^* \propto e^{-t/\tau},
\label{eq:2.4.1.14}\end{equation}
where $\tau$ is a relaxation parameter, dependent on stress value: 
\begin{equation}
\tau = 1/\ln(n^{*2}/\sigma)= 1\Big/\ln\left[\left({\textstyle\frac{1}{2}}+\sqrt{{\textstyle\frac{1}{4}}-\sigma}\right)^2\Big/\sigma\right] .
\label{eq:2.4.1.15}\end{equation}
At the critical stress, $\sigma = \sigma_c=\frac{1}{4}$, the argument of 
the logarithm is $1$ and apparently $\tau$ is infinite. 
As the critical stress is approached, 
for $\sigma \rightarrow \sigma_c$ 
\begin{equation}
\tau \simeq {\textstyle \frac{1}{4}}\;(\sigma_c-\sigma)^{-\theta}\hspace{5mm}\mbox{ with }\hspace{3mm} \theta = {\textstyle \frac{1}{2}}.
\label{eq:2.4.1.16}\end{equation}
 This divergence  clearly shows the character of the breaking 
 dynamics, i.e., it becomes very very slow at at the critical point.\\

\subsection{Universal behavior}
\begin{figure}[h]
\begin{center}
\includegraphics[width=8cm]{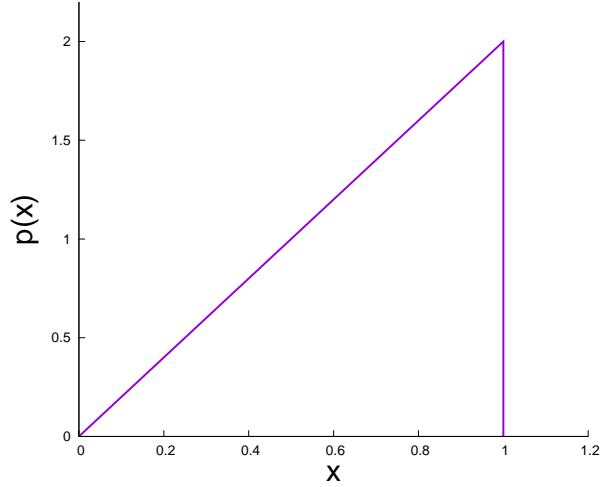}
\caption {\label{fig:inc-dist}
The linearly increasing fiber strength distribution.}
\end{center}
\end{figure}

The recursion relation and the fixed point solutions demonstrated the dynamic 
critical behavior for the uniform 
distribution of the 
breaking thresholds. Now the question arises -how general the results are? 
The {\it universality} of the cooperative breaking dynamics can be 
verified by considering a different 
distribution of fiber strengths.    
We are now going to examine the situation for a {\it linearly increasing 
distribution} (Figure \ref{fig:inc-dist}) within the
 interval $(0,1)$,
\begin{equation}
p(x) = \left\{\begin{array}{ll}
2x, & \hspace{5mm}0\leq x \leq 1\\
0& \hspace{5mm} x >1.
\end{array} \right.
\label{eq:2.4.3.1}\end{equation} 
From the force-stretch relationship, the average force per fiber is
\begin{equation}
F(x)/N = \left\{ \begin{array}{ll}
x(1-x^2)& \hspace{5mm} 0\leq x \leq 1\\
0       & \hspace{5mm} x>1,
\end{array}\right. .
\label{eq:2.4.3.2}\end{equation}
Therefore, the critical point is 
\begin{equation}
\sigma_c= \frac{2}{3\sqrt{3}}. 
\label{eq:2.4.3.3}\end{equation}
In this case the breaking dynamics can be written as a recursion relation:
\begin{equation}
n_{t+1}= 1-(\sigma/n_t)^2,
\label{eq:recur-inc}
\end{equation}
and the fixed-point equation is 
\begin{equation}
(n^*)^{3} -(n^*)^{2} +\sigma^{2} =0,
\label{eq:fixed-inc}\end{equation}
i.e., a cubic equation in $n^*$. 
Clearly, there are three solutions of $n^*$ for 
a value of $\sigma$. 
At the critical stress value, $\sigma_c=2/3\sqrt{3}$, the only 
acceptable solution of (\ref{eq:fixed-inc}) is 
\begin{equation}
n^*_c = {\textstyle \frac{2}{3}}.
\label{eq:2.4.3.10}\end{equation}

We want to investigate the breaking dynamics in the neighborhood of the 
critical point. Therefore, we insert $n=\frac{2}{3}+(n-n_c)$ into 
(\ref{eq:recur-inc}), with the result
\begin{equation}
{\textstyle \frac{4}{27}} - (n-n_c)^2-(n-n_c)^3 = \sigma^2 = ({\textstyle \frac{2}{3\sqrt{3}}}+\sigma-\sigma_c)^2={\textstyle \frac{4}{27}}+{\textstyle \frac{4}{3\sqrt{3}}} (\sigma - \sigma_c) + (\sigma-\sigma_c)^2.
\label{eq:2.4.3.11}\end{equation}
We get (to leading order)
\begin{equation}
(n-n_c)^2 = {\textstyle\frac{4}{3\sqrt{3}}}(\sigma_c-\sigma).
\label{eq:2.4.3.12}\end{equation}
Obviously, for $\sigma \leq \sigma_c$ the {\it order parameter} behaves as
\begin{equation}
n(\sigma)-n_c \propto (\sigma_c-\sigma)^{\beta}, \hspace{8mm} \beta = {\textstyle \frac{1}{2}},
\label{eq:order-linear}\end{equation}
in accordance with (\ref{eq:order-uni}). 
The {\it susceptibility} $\chi=-dn/d\sigma$ gives
\begin{equation}
\chi \propto (\sigma_c-\sigma)^{-\gamma}, \hspace{8mm} \gamma={\textstyle \frac{1}{2}}.
\label{eq:kappa-linear}\end{equation}

We can also discuss how the stable fixed point is approached from below.
From (\ref{eq:recur-inc}) one can write, around the fixed point
\begin{equation}
n_{t+1}-n^* = \frac{\sigma^2}{n^{*2}}-\frac{\sigma^2}{n_t^2} = \frac{\sigma^2}{n^{*2}n_t^2}\;(n_t^2-n^{*2} )\simeq (n_t-n^*)\;\frac{2\sigma^2}{n^{*3}}.
\label{eq:2.4.3.15}\end{equation}
The approach is clearly exponential, 
\begin{equation}
n_t-n^* \propto e^{-t/\tau} \hspace{8mm} \mbox{with } \hspace{4mm}\tau = \frac{1}{\ln (n^{*3}/2\sigma^2)}.
\label{eq:2.4.3.16}\end{equation}
The argument of the logarithm becomes $1$ exactly at the critical point, 
therefore 
$\tau$ diverges when the critical state is approached. The nature of such 
divergence assumes the same form, 
\begin{equation}
\tau \propto (\sigma_c-\sigma)^{-\theta}, \theta={\textstyle \frac{1}{2}},
\end{equation}
which is similar to the model with a uniform fiber strengt distribution, 
Eq. (\ref{eq:2.4.1.16}).

We can now conclude that, the ELS FBM with a linearly increasing 
fiber strength distribution  possesses the same critical power laws as 
the ELS FBM with a uniform fiber strength distribution. 
This confirms that the critical 
properties of the cooperative breaking dynamics are universal. 
A general treatment for verifying universality in 
ELS FBM can be found in Reference \cite{hhp15}.

\subsection{Two-sided critical divergence}

When fixed amount of load is applied on the system, the iterative 
breaking process ends with one of two 
possible end results. Either the whole bundle collapses, or an equilibrium 
situation is reached where intact fibers can hold/support the applied 
load/stress. Thus, the final fate of the bundle 
depends on whether the external stress $\sigma$ on the bundle is postcritical  
($\sigma > \sigma_c$), precritical ($\sigma<\sigma_c$), or critical 
($\sigma=\sigma_c$). It is interesting to know how the breaking dynamics is
approaching the critical point (failure point) from below (pre-critical) and 
above (post-critical) stress values. 

In case of  uniform fiber strength distribution when the external stress 
approaches the critical value $\sigma_c=1/4$ from a 
higher value, i.e., in the post-critical region, the number 
of necessary iterations needed for the whole system to break increases as the 
critical point is approached. 
Close to the critical point, number of iterations shows a square-root 
divergence \cite{ph07}:
\begin{equation}
t_f \simeq {\textstyle \frac{1}{2}}\pi (\sigma - \sigma_c)^{-1/2}. 
\label{eq:2.4.4.15}\end{equation}
\begin{figure}[h]
\begin{center}
\includegraphics[width=8cm]{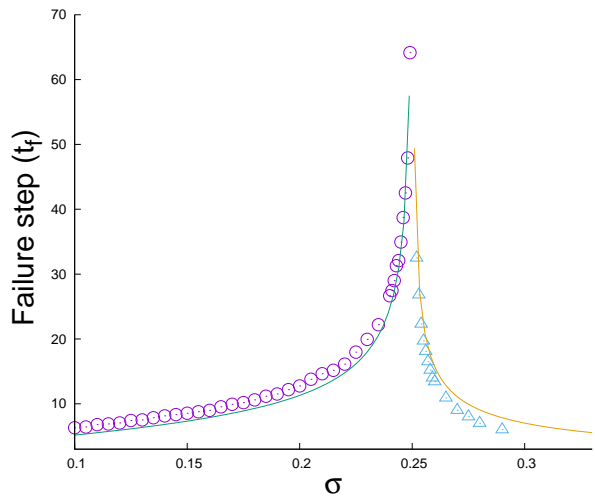}
\caption {\label{fig:post-crit}
Post-critical and pre-critical relaxation: Numerical data are for a bundle 
with $N=10^6$ fibers having uniform threshold distribution and averages are 
taken over $10^5$ samples. Lines are showing theoretical estimates.} 
\end{center}
\end{figure}

Similarly, in the pre-critical region, when the external stress approaches the 
critical value $\sigma_c=1/4$ from below, the number of iterations has again a 
square root divergence \cite{ph07} (for uniform distribution) close to the 
critical point:
\begin{equation}
t_f = {\textstyle \frac{1}{4}}\;\ln (N)\,(\sigma_c-\sigma)^{-1/2}.
\label{eq:2.4.5.15}\end{equation}
The only difference is that, in pre-critical case, the amplitude of the square 
root divergence has a system-size-dependence which is absent in post-critical 
case.  

We can conclude that in ELS FBM, the breaking dynamics shows a two-sided 
critical divergence in terms of the number of iteration steps 
needed to reach critical 
point from below (pre-critical) and above (post-critical) 
(Figure \ref{fig:post-crit}).
The theoretical details of the exact solutions can be found in References \cite{ph07,hhp15}. 

\subsection{Avalanche dynamics with fixed amount loading}
The number of fibers ($S$) breaking between two successive stable conditions of the fiber bundle is called an avalanche. The distribution
of the avalanche sizes $P(S)$ shows a power-law tail for the large $S$ limit \cite{HH}, which is a sign of the criticality discussed above. This
is experimentally widely observed for driven disordered systems in general \cite{brc15} and (quasi-brittle/ductile) fracture in particular.  While 
the details of the avalanche dynamics seen in the fiber bundle model with quasi-static load increase has been discussed elsewhere in this special
issue \cite{jonas}, here we briefly describe the avalanche dynamics for fixed amount load increase i.e., when the system is in a stable condition, a 
fixed amount of load $\delta$ is added, which restarts the dynamics. 
As before, the number of fibers breaking until the system reaches the next 
stable state constitutes an avalanche. Clearly, this type of avalanche is 
a result of the cooperative breaking dynamics and it is not arising 
due to any fluctuations in stress levels or in fiber strength distribution.
We will describe below how to calculate theoretically the distribution of 
such avalanches:

The load curve, in terms of the threshold values, can be written as
\begin{equation}
F(x)=Nx(1-x)
\label{load-curve-uniform}
\end{equation} 
for uniform threshold distribution in $(0,1)$ (see Eq. (\ref{eq:load-curve})). The load increases between $0$ and $N/4$ with increment $\delta$. Therefore,
the values of the load are $m\delta$, with $m=0,1,2, \dots, N/4\delta$. The threshold value for load $m\delta$ can be obtained from Eq. (\ref{load-curve-uniform})
as
\begin{equation}
x_m=\frac{1}{2}(1-\sqrt{1-4m\delta/N}).
\end{equation}
The average number of fibers broken due to the increase of load from $m\delta$ to $(m+1)\delta$ is
\begin{equation}
S=N\frac{dx_m}{dm}=\frac{\delta}{\sqrt{1-4m\delta/N}}.
\end{equation}
The number of avalanches of size between $S$ and $S+dS$ is obtained from the corresponding interval of the variable $m$ i.e., $P(S)dS=dm$. From the equation
above, we have
\begin{equation}
\frac{dS}{dm}=2S^3/(N\delta).
\end{equation}
Therefore, the avalanche size distribution is given by 
\begin{equation}
P(S)=\frac{dm}{dS}=\frac{1}{2}N\delta S^{-3}, \mbox{for} S \ge \delta.
\label{avalanchesize-discrete}
\end{equation}
Indeed, it is possible to show \cite{hhp15} that for an arbitrary threshold distribution $p(x)$, the large $S$ asymptotic limits of
the avalanche size distribution is
\begin{equation}
P(S)\sim CS^{-3},
\label{avalanche_eq}
\end{equation} 
with $C=N\delta\frac{p(x_c)^2}{2p(x)+x_cp^{\prime}(x_c)}$,
with the mild assumption that the load curve has a generic parabolic form with a critical point.

\section{Noise-induced failure in FBM}
So far we have discussed the classical stress-induced failure of fibers 
without the presence of noise. A noise-induced failure scheme for 
fiber bundle model can be formulated \cite{Roux,pc03,pcc13} for which the cooperative failure
dynamics can be solved analytically. 

 As in the previous sections, we consider  a bundle of $N$ parallel 
 fibers clamped between two rigid bars. A load ior force ($F=\sigma N$) is 
 applied on 
 the bundle. The fibers have different strength thresholds ($x$) 
 and there is a critical strength $\sigma_c$ \cite{RMP} for the whole bundle, 
 so that the bundle 
does not fail completely for stress $\sigma \le \sigma_c$, but it fails 
immediately for $\sigma>\sigma_c$. 
Now we introduce noise ($T$) in the system 
and  assume that each fiber having 
strength $x_i$ has
a finite probability $P_f(\sigma,T)$ of failure at any stress
$\sigma$ induced by a noise $T$:
\begin{equation}
\label{july31-1}
P_f(\sigma ,T)= \left\{ \begin{array}{cc}
C\exp \left[ -\frac{1}{T}\left( \frac{x_i}{\sigma }-1\right) \right] , & 0\leq \sigma \leq x_i\\
1, & \sigma >x_i
\end{array}\right .
\end{equation}
Here $C$ is a prefactor. $P_f(\sigma,T)$ increases as $T$ increases and for a 
fixed value of 
$T$ and $\sigma_c$, as we increase $\sigma$, the bundle breaks more rapidly. 
The motivation behind Eq. (\ref{july31-1}) comes from the time-dependent 
behavior 
or so-called creep behavior of materials, observed in real systems 
\cite{Lawn,hhp15}. It is obvious that strength of elements/fibers degrades in 
time due to external influences like moisture, temperature etc. 

Such a noise-induced failure scheme will produce two different failure regimes 
depending on the stress and noise levels -continuous breaking regime and 
intermittent breaking regime. In the continuous breaking regime we can 
calculate 
the failure time (step) as a function of stress and noise values. However, in 
the intermittent breaking regime one can define the waiting time between two 
consecutive 
failure phases.

\begin{figure}
\includegraphics[width=8cm]{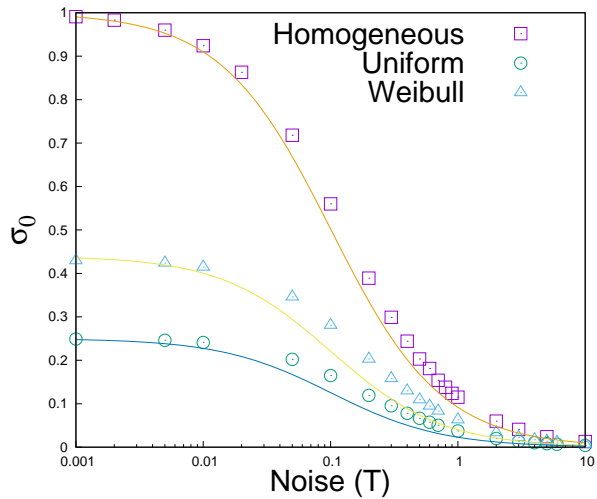}
\caption{\label{fig:fg1}
Phase boundary ($\sigma_0$ vs. $T$ plot) for three different type of fiber 
strength distributions with $N=20000$. Data points are simulation results and 
	solid lines are analytic estimates (Eq. \ref{noise_ph_boundary}) based on mean-field arguments.}
\end{figure}

The phase boundary can be determined through a mean-field argument that at 
$\sigma = \sigma_0$, at least one 
fiber must break to trigger the continuous fracturing process. After this 
single failure the applied load has to be redistributed on the intact fibers 
(due to ELS) and the 
effective stress will surely increase (more than $\sigma_0$), which in turn 
enhances failure probability for all the intact fibers. 
Following this logic, in case of homogeneous bundle where all the fibers have 
identical strength $x_i=1$ (and $\sigma_c=1$), at the phase boundary
$N P_f(\sigma_0,T) \ge 1$
giving
\begin{equation}
N \exp\left[-\frac{1}{T}\left(\frac{1}{\sigma_0}-1\right)\right] \ge 1
\end{equation}
which finally gives
\begin{equation}
\sigma_0 \ge \frac{1}{1-T log(1/N)}.
\label{noise_ph_boundary}
\end{equation}
In absence of noise, when $T=0$, the above equation gives $\sigma_0=1=\sigma_c$, which is consistent with the 
static FBM results \cite{RMP}.  
This analytic estimate overlaps with the data obtained from simulation 
(Fig.~\ref{fig:fg1}). It shows that the continuous 
and intermittent fracturing regimes are separated by a well defined phase 
boundary which depends on both, the stress level and the noise level 
\cite{pcc13}. 

In case of  heterogeneous FBM where fibers have different 
strength thresholds, 
keeping in mind that in absence of noise $T$, we should always get
$\sigma_0=\sigma_c$,  one can  make a conjecture that 
\begin{equation}
\sigma_0 \ge \frac{\sigma_c}{1-T log(1/N)}.
\end{equation}
The numerical data for the heterogeneous cases (Fig. \ref{fig:fg1}) having 
uniform and Weibull type fiber strength distributions supports the 
conjecture well \cite{pcc13}.

Identification of such a phase boundary has important consequences in 
material-fracturing and in other similar fracture-breakdown phenomena.   
During material/rock fracturing, acoustic emission (AE) measurements
can record the burst or avalanche events in terms of AE amplitude and AE 
energy \cite{AE}.
Therefore, AE data could reveal the correct rupture-phase of a material body 
under stress.  Once a system enters into continuous 
rupture phase the system-collapse must be imminent. Thus 
identification of rupture phase can guide us to visualize the final fate of a 
system. It can also help to 
stop system collapse, if it is possible to withdraw external stress in time 
-before the system enters into continuous rupture phase.

We will now  discuss cooperative dynamics in both these regimes 
in the following sub-sections.     

\subsection{Continuous breaking regime}
 In the continuous breaking regime one can describe the breaking dynamics 
 in a FBM through a recursion relation \cite{pc03}. Let us consider a 
 homogeneous bundle 
 having $N$ fibers with exactly same strength thresholds $1$; therefore 
 critical (or failure) strength of the bundle is $\sigma_c=1$. Now we consider 
 a noise-induced failure probability for breaking of each fiber in the continuous regime: 

\begin{equation}
\label{prob-homo}
P_f(\sigma ,T)= \left\{ \begin{array}{cc}
	\frac{\sigma}{\sigma_c}\exp \left[ -\frac{1}{T}\left( \frac{x_i}{\sigma }-1\right) \right] , & 0\leq \sigma \leq x_i\\
1, & \sigma >x_i
\end{array}\right .
\end{equation}
As all the fibers are identical, $x_i=1=\sigma_c$. The prefactor is a function 
of stress level $\sigma$ and this is a careful choice to get a solution of 
the recursive dynamics, which we will describe below.

We denote the fraction of total fibers that remain intact at time (step) $t$ by $n_t$ and the breaking dynamics can be written as

\begin{equation}
	n_{t+1}=n_t\left[1-P_f\left(\frac{\sigma}{n_t},T\right)\right].
\label{noise_rec}
\end{equation}
In the continuum limit the above recursion relation can be presented in a 
differential form 
\begin{equation}
	-\frac{dn}{dt}=\frac{\sigma}{\sigma_c} exp\left[-\frac{1}{T}\left(\frac{\sigma_c}{\sigma}n-1\right)\right].
\end{equation}
giving the failure time 
\begin{equation}
	t_f=T exp\left(-\frac{1}{T}\right)\left[exp\left(\frac{\sigma_c}{\sigma T}\right)-1\right].
\label{noise_tf}
\end{equation}

\begin {figure}[t]
\includegraphics[width=8cm]{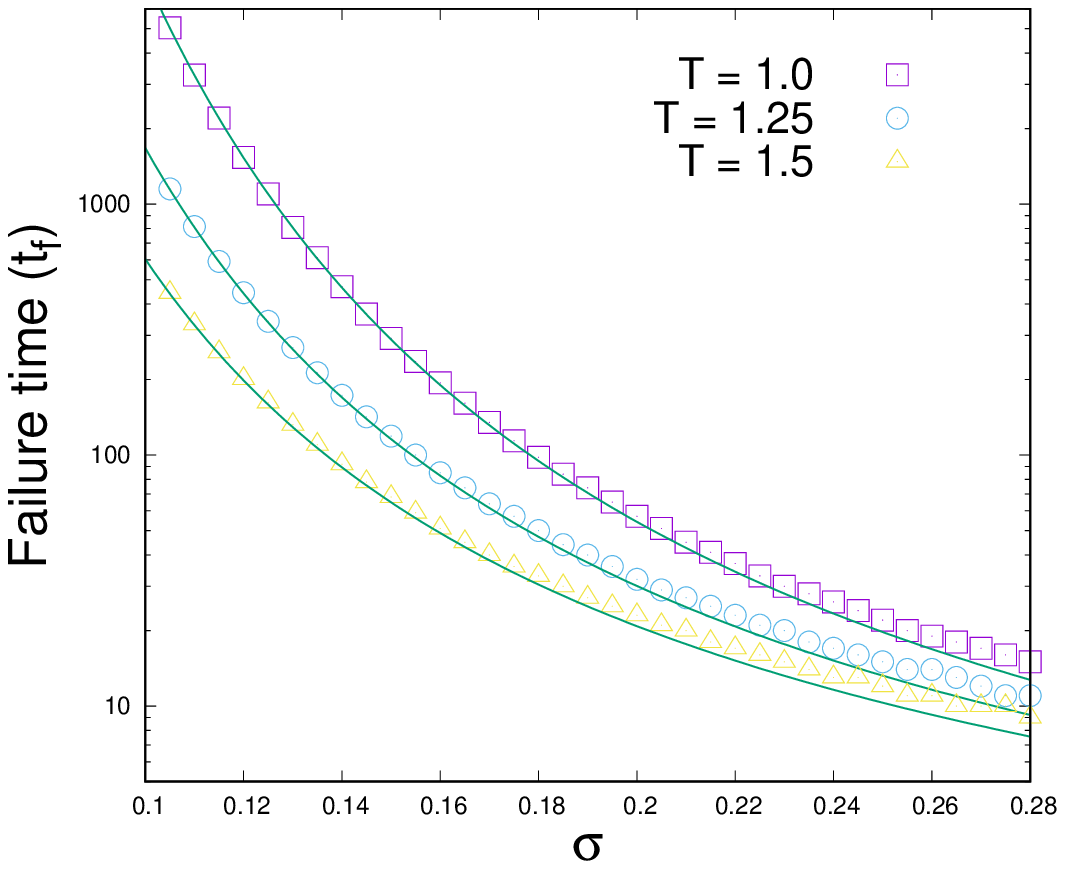}
\includegraphics[width=8cm]{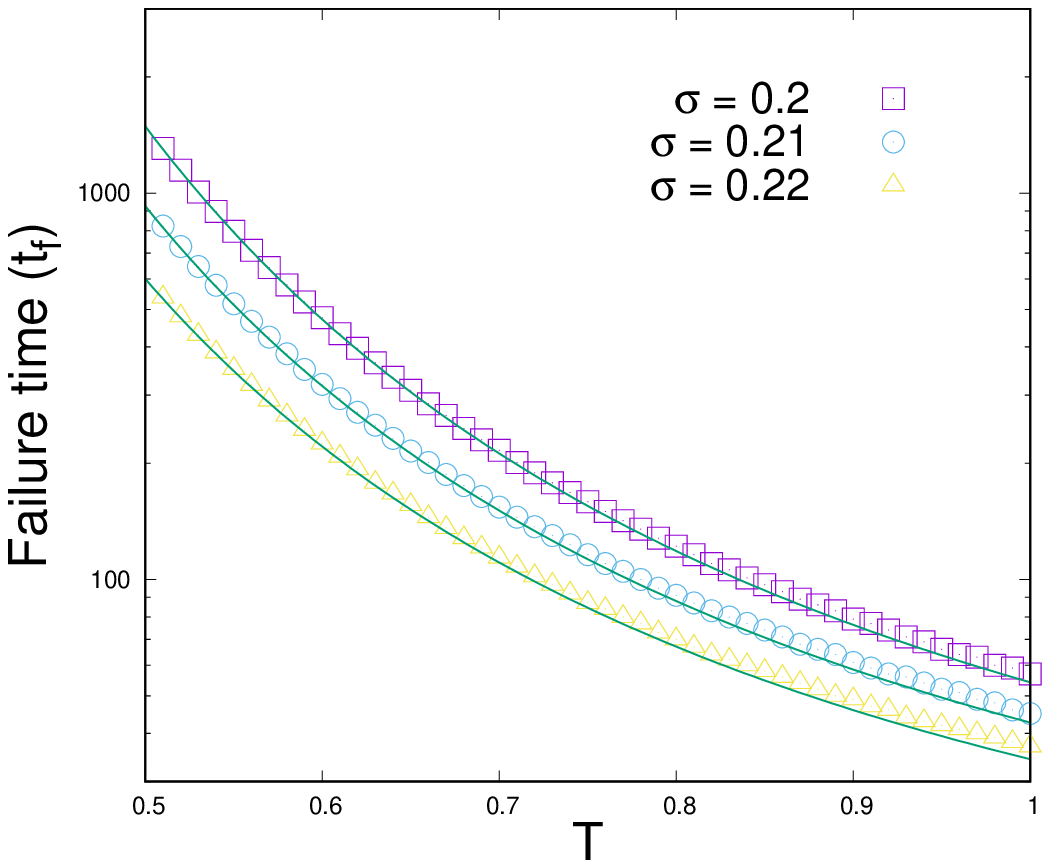}
\caption{Failure time vs. $\sigma$ (left) and vs. $T$ (right) for a homogeneous bundle having identical fibers with strength $1$ ($\sigma_c=1$ as well). The data are for simulations over a single realization with system size $N =1000000$ and the solid lines are the theoretical estimates following Eq. (\ref{noise_tf}).} 
\label{fig-homo}
\end {figure}

The simulation result shows (Fig.\ref{fig-homo}) exact 
agreement with this 
theoretical estimate. 

\begin {figure}[t]
\includegraphics[width=8cm]{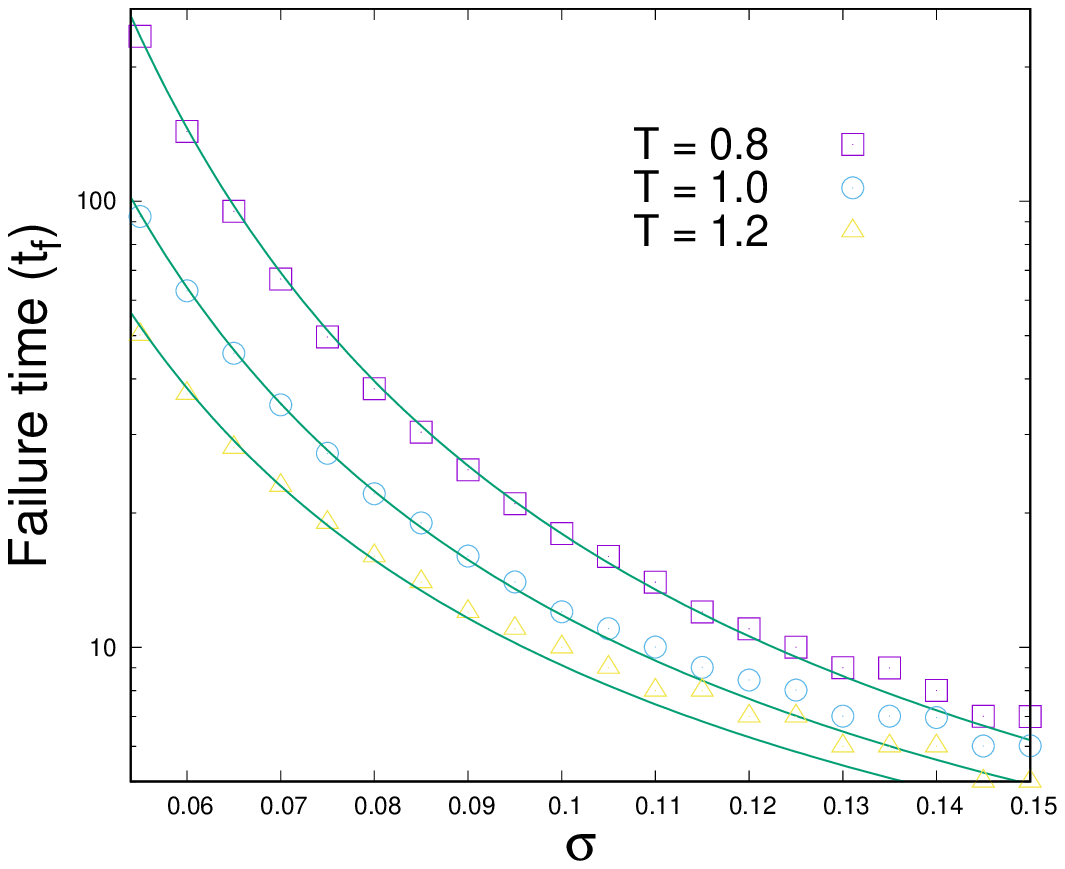}
\includegraphics[width=8cm]{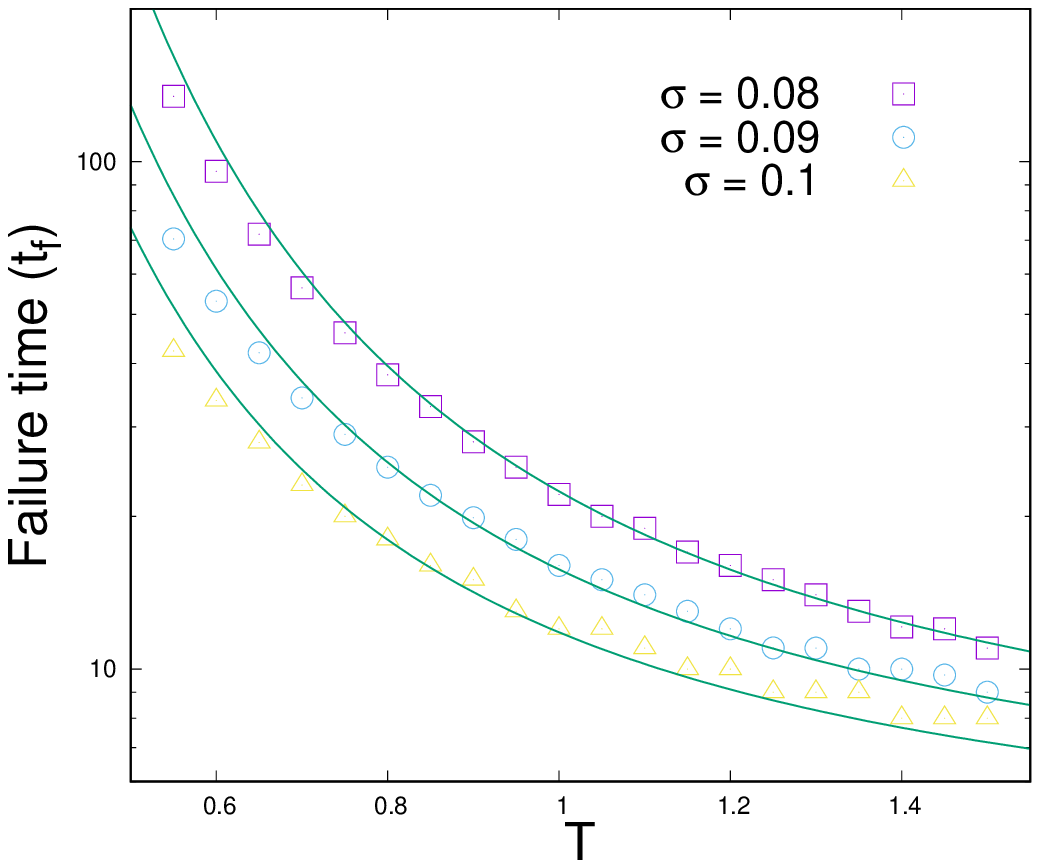}
\caption{Failure time vs. $\sigma$ (left) and vs. $T$ (right) for  bundles having uniform strength distributions. The data are for simulations over 1000 realizations with system size $N =10^6$ and the solid lines are the theoretical estimates following Eq. (\ref{noise_tf_het}).} 
\label{fig-hetero}
\end {figure}

In case of heterogeneous bundles where fibers have distributed strengths the 
failure times seem to follow another form \cite{pc03}:   
\begin{equation}
	t_f=T exp\left(-\frac{1}{T}\right)\left[exp\left(\frac{\sigma_c}{\sigma T}+\frac{1}{T}\right)-1\right].
\label{noise_tf_het}
\end{equation}
This form was obtained through trial and error approach. It is 
extremely difficult (as of now) to write the recursion relation for
noise-induced failure dynamics in case of heterogeneous systems. 
The simulation results have been compared with the formula above and the 
agreement (Fig.\ref{fig-hetero}) is quite satisfactory \cite{pc03}.

\subsection{Intermittent regime}
\begin{figure}[h]
\begin{center}
\includegraphics [width=8cm]{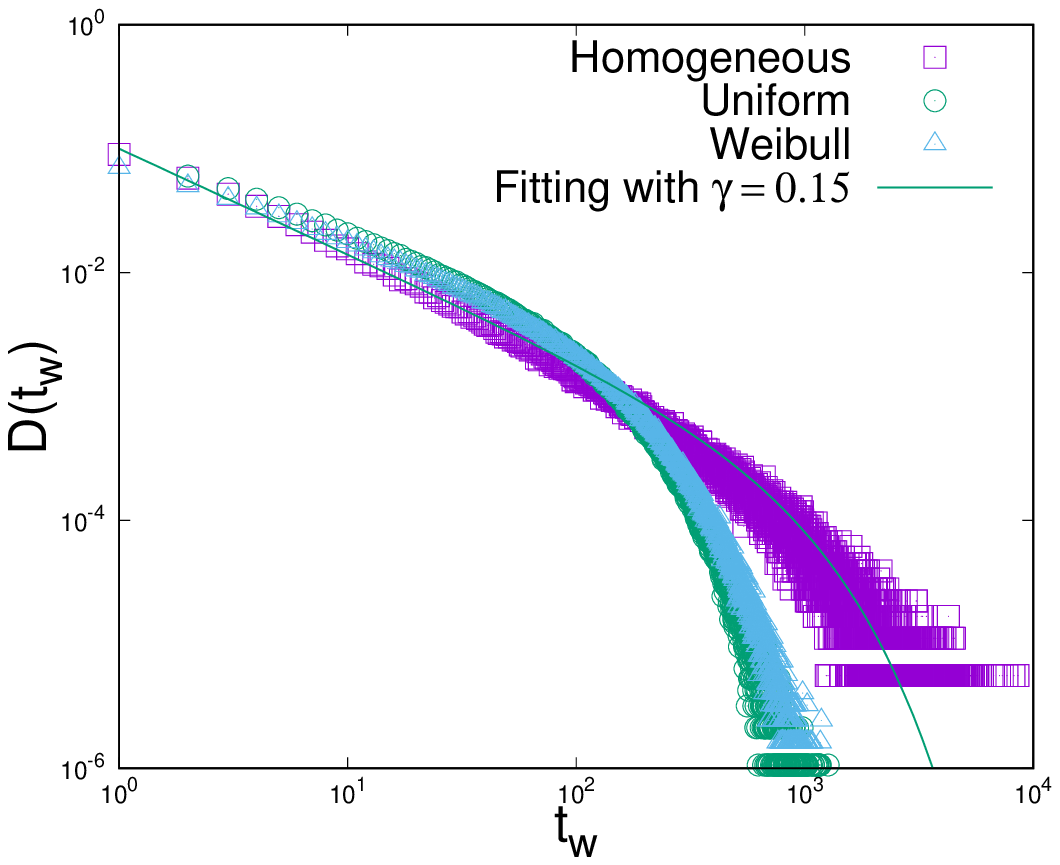}
\includegraphics [width=8cm]{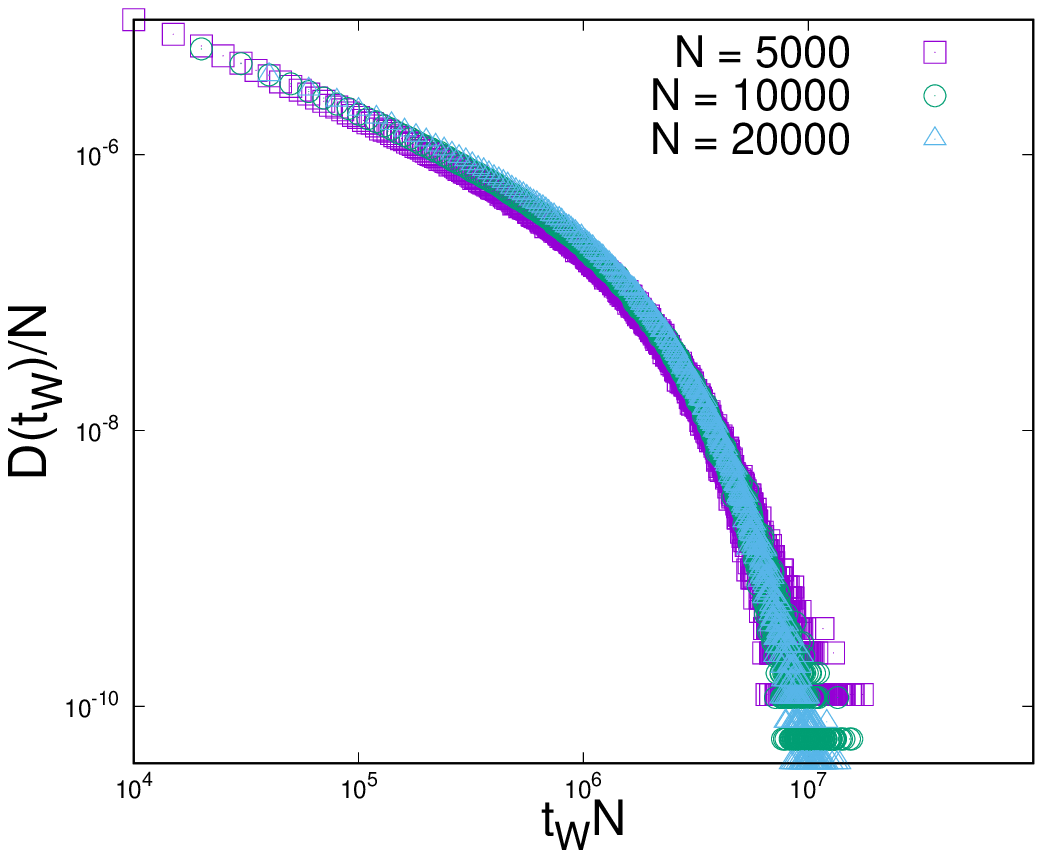}
\end{center}
\caption{ Left: The simulation results for the waiting time distributions for 
 three different type of fiber strength distributions with $N=20000$. 
All the curves can be fitted  with the Gamma form $exp(-t_W/a)/t_W^{1-\gamma}$, 
where  $\gamma = 0.15$ for homogeneous case and  $\gamma = 0.26$ for 
uniform and Weibull distributions. 
Right: we show the data collapse of the waiting time distributions with 
system sizes for uniform fiber strength distribution.} 
\label{fig:fg2}
\end{figure}

As we discussed before, in the intermittent fracturing phase, simultaneous 
breaking events (avalanches) are
separated by waiting times ($t_W$) of different magnitude.
The waiting time 
distribution can be fitted with a Gamma distribution \cite{pcc13} for both 
homogeneous and heterogeneous bundles
\begin{equation}
D(t_W) \propto \exp(-t_W/a)/t_W^{1-\gamma}
\label{gamma_dist}
\end{equation}
where $\gamma = 0.15$ for homogeneous case and  $\gamma = 0.26$ for 
heterogeneous cases (Fig.~\ref{fig:fg2}). 
Here $a$ is a measure of the extent of the power law
regime and it seems that the 
power law exponent does not change with the variation of 
$\sigma$, $T$ and $N$ \cite{pcc13}. 

In the waiting time distributions, the power law part dominates  
for small $t_W$ values and exponential law dominates for bigger $t_W$ values.  
The inherent global load sharing nature is responsible 
for  the power law part of the Gamma distribution, as power law usually 
comes from a 
long range cooperative mechanism \cite{CB97,HR90,HH}. The exponential part of 
the Gamma distribution is contributed by the noise induced 
failure factor $P_f(\sigma,T)$. For large $t_W$ values one can eventually treat
the failures to be independent. If $P$ indicates the noise induced failure 
 probability within $t_W$ then 
the probability
$D(t_W)=A(1-P)^{t_WN}\sim exp(-Pt_WN),$
where $A$ is a constant. The normalization of $D(t_W)$ requires $A\sim N.$ 
Though for smaller values of $t_W$, one can not ignore the correlations between
 successive failures (responsible for the power law part in $D(t_W)$), the 
exponential scaling behavior for $D(t_W)$ can be easily obtained from the 
above. As shown in the inset of Fig.~\ref{fig:fg2}, the plot of  $D(t_W)/N$ 
against $t_W N$ gives good data collapse for different $N$ values. Such a data 
collapse indicates the robustness of the Gamma function form. It is not clear 
yet whether the Gamma type distribution is a direct consequence of the 
failure probability function (Eq. \ref{july31-1}). It needs more 
investigations with various other types of possibilities for 
Eq. (\ref{july31-1}).

\begin{figure}
\includegraphics[width=8cm]{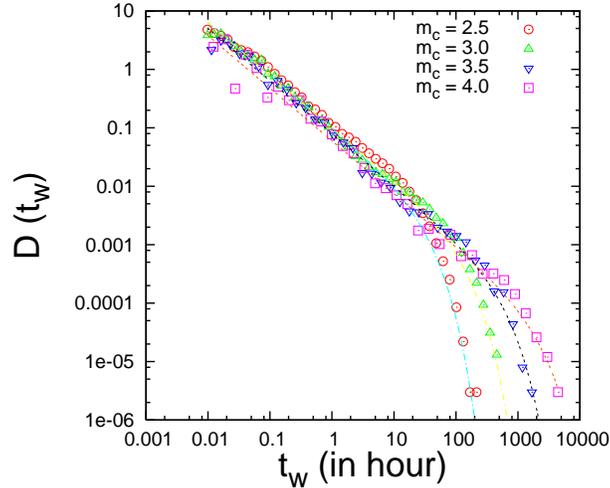}
\caption{Gamma-fitting (dotted lines) to the waiting time distributions in 
California catalog (1984-2002).} 
\label{fig:fg3}
\end{figure}

Apparently, the modeling scheme for noise-induced rupture process is not 
limited to any particular system, rather it is a general approach and perhaps 
it can model more complex situations like rupture driven earthquakes. 
In literature we can find evidences of  stress-localization around 
fracture/fault lines in a active seismic-zone. Also, there are 
several factors that can help rupture evolution, like friction, plasticity, 
fluid migration, spatial heterogeneities, chemical reactions etc.  
To some extent, such stress redistribution/localization can be taken into 
account through a proper load sharing scheme and a noise term ($T$) can 
in principle represent the combined effect of all other factors.    

To compare the waiting time results of the model system with real data, 
the California earthquake catalog from 1984 to 2002 \cite{catalog} has been 
analyzed \cite{pcc13} to study the statistics of 
waiting times \cite{Corral06,Corral03,Bak} between earthquake events. 
First, a cutoff ($m_c$) has been set in 
the earthquake magnitude - so that all earthquake events above this cutoff 
magnitude will be considered for the analysis. 
The distribution of 
waiting times shows similar variation for different cutoff values. It seems
\cite{pcc13} waiting time distributions for 
all the data sets follow a Gamma distribution \cite{Corral06}:
\begin{equation} 
D(t_W)\propto exp(-t_W/a)/t_W^{(1-\gamma)};
\label{gamma_fn}
\end{equation} 
with same $\gamma$ ($\simeq$ 0.1) and different $a$ values for different cutoff levels:
$a=30, 120, 500, 2000$ respectively for $m_c=2.5, 3.0, 3.5, 4.0$ (see Fig.~\ref{fig:fg3} ).  

The similarities in waiting time statistics and scaling forms suggest 
that slowly driven (noise-induced) fracturing process and earthquake 
dynamics (stick-slip mechanism) perhaps have some common origin.

\section{Interface propagation in the fiber bundles: Self-organization and depinning transition}
So far we have considered FBM versions that are globally loaded i.e., all the fibers in the system are
loaded equally from the initial time, and the load remains equal on each surviving fiber, given that the 
load sharing is equal. This necessarily imply that the damage or failures in the system could occur at
any point; indeed there is no notion of distance in this form of the model. 

However, in fracture dynamics, particularly in the mode-1 variant of it, a front could propagate in the
direction transverse to that of the loading. A fracture front necessarily implies damage localization within a region 
with a lower dimension than that of the system i.e., a front-line in two dimensions or a front surface in a three
dimensional system. Indeed, driven front propagation through a disordered medium is not limited to fracture, but also
happens in vortex lines in superconductors \cite{vortex}, magnetic domain walls in magnetic materials with impurities \cite{domain_wall},
contact line dynamics in wetting \cite{c_line} and so on.  
  
\begin{figure}[tb]
\centering \includegraphics[width=9cm]{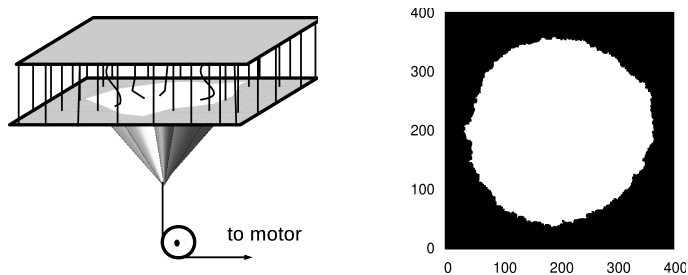}
   \caption{A schematic representation of locally loaded fiber bundle model and the resulting interface propagation. From \cite{bc13}.}
\label{schematic_interface}
\end{figure}

In the context of FBM, it is possible to capture the dynamics of a front propagating through a disordered medium 
by considering a localized loading of the system (when the fibers are arranged in a square lattice and the load is applied at an 
arbitrarily chosen central site; see Fig. \ref{schematic_interface})
 in dimension higher than one (in one dimension 
the damage interface is a point and hence cannot increase).
The external load is increased at a low and constant rate (maintaining the separation of time scales between applied loading rate and redistribution process) \cite{bc13}.
Initially the system is not loaded anywhere except for the one fiber at an arbitrarily chosen central site. As the external load increased beyond the 
failure threshold of that central
fiber, it breaks and the load carried by that fiber is redistributed among the fibers that are in the damage boundary (in the beginning just the four nearest neighbors). 
  Therefore, the fibers that are 
newly exposed to the load after an avalanche, carry a lower load
compared to those accumulating loads from the earlier avalanches. 
This process keeps
a compact structure of the cluster of the broken fibers. 
The localized nature of the load redistribution is justified from the fact that the newly 
exposed fibers are further away from the point of loading and therefore carry 
a smaller fraction of the load at the original central site. 

As the damage perimeter increases, so does the number of fibers on that perimeter. This implies that for an avalanche, the load per fiber
will decrease along the damage boundary. But due to further increase in the load, this value will subsequently increase, initiating another
avalanche. In the steady state, the load per fiber value will fluctuate around a constant and the system is said to have reached a self-organized state.
In this state, the failure of fibers  in 
the process of avalanches have a scale free size distribution, which suggests that it is a 
self-organized critical (SOC) state (where external drive and dissipation balance and the critical point becomes an attractive fixed point \cite{brc15}). 

The steady state value of the load per fiber and the corresponding avalanche size distribution can be calculated for a variant of
this model where the load redistribution is uniform along the entire damage boundary i.e. every fiber along the damage boundary gets the same fraction of load in a redistribution process. We discuss this for the Weibull distribution below, but this is true for other distributions as well.

The Weibull distribution in its general form can be written as
\begin{equation}
W_{\alpha,\beta}(x)=\alpha \beta x^{\alpha-1}e^{-\beta x^{\alpha}},
\end{equation}
 where $\alpha$ and $\beta$
are the two parameters. We can consider the particular case when $\alpha=2$ and $\beta=1$. The failure threshold of a fiber
 is greater than $x$ with a probability that is proportional to $\int\limits_{x}^{\infty}x^{\prime}e^{-{x^{\prime}}^2}dx^{\prime}\sim e^{-x^2}$.
Given that the probability density function for force is uniform, the probability of a fiber having load
between $x$ and $x+dx$ is $e^{-x^2}P(x)dx$, with $P(x)=c$ (unnormalized). The normalization gives $c\int\limits_0^{\infty}e^{-x^2}dx=1$ 
implying $c=\frac{2}{\sqrt{\pi}}$. Hence the normalized probability density function for the load on the surviving fibers
is
\begin{equation}
D_{\sigma}(x)=\frac{2}{\sqrt{\pi}}e^{-x^2}.
\label{wei-force}
\end{equation}
 Similarly, the probability that the load is lower than $x$
is proportional to $x$. Using the form for threshold distribution ($\sim xe^{-x^2}$), the probability density
function for threshold distribution of the survived fibers becomes 
\begin{equation}
D_{th}(x)=\frac{4}{\sqrt{\pi}}x^2e^{-x^2}.
\label{wei-th}
\end{equation}
 Both
of these functions are in good agreement with numerical simulations. 
Also the saturation value of the average load per fiber
can be calculated  as
\begin{equation}
 \int\limits_0^{\infty}xD_{\sigma}(x)dx=\frac{2}{\sqrt{\pi}}\int\limits_0^{\infty}xe^{-x^2}dx=\frac{1}{\sqrt{\pi}},
\end{equation}
which is again in good agrees with simulations.

\begin{figure}[tb]
\centering \includegraphics[width=9cm]{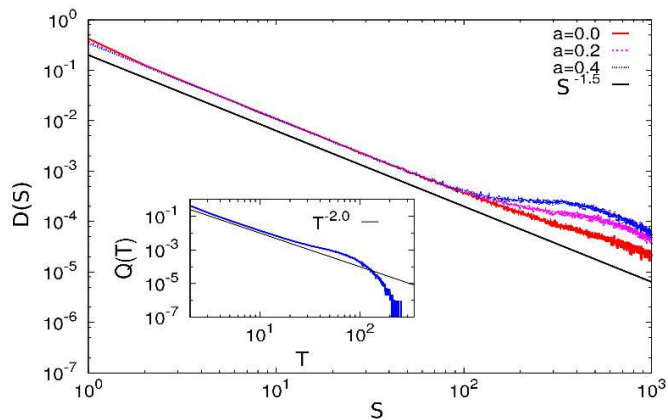}
   \caption{ The avalanche size distributions are plotted for zero and finite lower cut-offs
for Model II. 
The distribution function is a power-law with exponent value $1.50\pm 0.01$, which
is also our estimate from scaling arguments. Inset: The distribution of avalanche duration is plotted for Model II. 
This also shows a power-law decay
with exponent value $2.00\pm 0.01$. From \cite{bc13}.}
\label{avalanche-comb}
\end{figure}

The size distribution of avalanches is a power law with the exponent value
close to $3/2$ (see Fig. \ref{avalanche-comb}), which is in agreement with the scaling prediction of avalanche 
size distributions in SOC models for mean field.
 The distribution of the avalanche duration i.e., the number of redistribution steps for an avalanche, is a 
power-law with exponent value close to $2.00\pm 0.01$.
which is again in agreement with
scaling  predictions of SOC models in mean field. 

For estimating the avalanche size exponent, it can be assumed that the average load per fiber on the 
damage boundary has a 
distribution which is Gaussian around its mean: $P(\sigma )\sim e^{-(\sigma-\sigma_c)^2/{\delta \sigma}}$. Hence, from a
dimensional analysis,  mean-squared
fluctuation $\delta \sigma\sim (\sigma-\sigma_c)^2$ .
Also the avalanche size $S$ scales as $(\delta \sigma)^{-1}$ since it may be viewed as the number of broken fibers
after a load increase of $\delta \sigma$. This gives
\begin{equation}
(\sigma-\sigma_c)\sim S^{-1/2}.
\label{sig1}
\end{equation} 
The probability of an avalanche being of the size between
$S$ and $S+dS$ is $D(S)dS$. Now, the deviation from the critical point scales \cite{RMP}
with the cumulative size of all avalanches upto that point;
giving $(\sigma-\sigma_c)\sim \int\limits_{S}^{\infty}D(S)dS$. If we take $D(S)\sim S^{-\gamma}$,
then 
\begin{equation}
(\sigma-\sigma_c)\sim S^{1-\gamma}.
\label{sig2}
\end{equation}  
Comparing Eqs. (\ref{sig1}) and (\ref{sig2}) therefore we have $\gamma=\frac{3}{2}$. So the probability density function for
the avalanche size becomes $D(S)\sim S^{-3/2}$, which fits well with simulation results (Fig. \ref{avalanche-comb}).

\section{Some related works on the dynamics of FBM}

In this section we would like to bring attention to some related works on 
the dynamics of FBM which, we believe,
may be regarded as essential reading in this field. 

As we have discussed in detail in the earlier sections, there has been considerable progress in characterizing the failure dynamics in the fiber bundle model through
tools describing critical phenomena. One crucial step towards that direction is to identify the universality class of the model. That often needs 
a coarse grained description of the model, writing down the free energy form suited for the dynamics and then identifying the symmetries and 
consequently the universality class. One such step was done in Ref. \cite{hendrk} by writing down a mesoscopic description of the ELS-FBM.
specifically, writing the time evolution of the order parameter $n(\sigma)-n_c=\eta$ and the driving field (stress increase) as $J=\sigma_c-\sigma$, 
the dynamics is described by
\begin{equation}
\frac{\partial \eta}{\partial t}=-\eta^2+J.
\end{equation}
 Writing in terms of the density of intact fibers $n$, 
\begin{equation}
\frac{\partial n}{\partial t}=\lambda n(1-n)-\sigma,
\end{equation}
with $\lambda=1$. This equation has a particle-hole symmetry for zero external field $\sigma=0$, hence generally expected to be in the CDP
or compact domain growth universality class of non-equilibrium phase transition \cite{cdp}.
Although done for the ELS version, this approach of relating fiber bundle model dynamics to nonequilibrium
critical phenomena through a Langevin equation could provide useful insights in more realistic versions. 

Among other attempts to relate fracture and in particular FBM dynamics with different universality classes, 
a relatively less explored route is that of the hydrodynamics of turbulence. The analogy between the velocity fluctuation in turbulence 
and surface roughness due to fracture have been explored before (see e.g., \cite{basu18}). However, given that FBM is able to provide a
reasonably consistent picture for fracture dynamics, its association with hydrodynamics of fracture is a crucial
question. In Ref. \cite{bc20} the relation between the Kolmogorov energy dispersion in turbulence and
avalanche dynamics in the FBM was explored. Specifically, the vortex lines in a fully developed
turbulence can be mapped to Self Avoiding Walk (SAW) picture of polymers \cite{huang}. Then following
Flory's theory \cite{flory}, the Kolmogorov energy dispersion becomes 
\begin{equation}
E_q \sim q^{-1/\nu_F},
\label{kol_eq}
\end{equation}
where $q$ is the wave number, $\nu_F$ is the Flory exponent and $d$ is the spatial dimension. 
Then drawing the parallel with the energy dispersion in avalanche dynamics in the FBM (see Eq. (\ref{avalanche_eq})),
we get $E_q\sim q^{-d/3}$ for the mean field case (i.e. $d=d_u$, the upper critical dimension). By taking 
$d_u=6$, which is consistent for the FBM \cite{skh15}, we get back the Flory mean field result $E_q\sim q^{-2}$. 
In parallel, by taking the correlation length as inverse of the wave number $q$, and using finite size 
scaling arguments, one can show that $\nu d=2/3$ in the mean field limit, where $\nu$ is the correlation length
exponent. Again using $d=6$ as the upper critical dimension, one gets $\nu=1/4$. 

It may be noted that there is also a gratifying
   consistency in the main results discussed above. In the
   ELS FBM, the
   critical exponents $\beta$, $\gamma$ and $\nu$ for the order
   parameter, breakdown susceptibility and correlation
   length respectively satisfy the Rushbrooke scaling
   relation (incorporating the hyperscaling relation) \cite{skma}:
   $2\beta + \gamma = d\nu$, with $\beta = 1/2  = \gamma$ and with the value of the upper  critical
   dimension $d = 6$ and $\nu = 1/4$.

Given that the fiber bundle is essentially an ensemble of discrete elements having finite failure thresholds, 
under the condition of conserved load, it can serve as a generic model for intermittent progress towards catastrophic failure
in a wide variety of systems. Such systems can be roads carrying traffic, power grids to redundant computer circuitry. In several
of such cases, the load redistribution following the failure of an individual element (say, traffic jam along one road, 
failure of one power station etc.), is controllable to some extent - a freedom lacking in the case of stressed disordered solids. 
Under such circumstances, it is useful to ask the question how the total load carrying capacity of the system could be maximized
by a suitable load redistribution rule \cite{bs15}. 

It is rather straightforward to establish that the maximum limit of $\sigma_c$ would be achieved, when the maximum number of 
fibers carry load to their fullest capacity. For a uniform distribution of the failure thresholds in (0,1), it is possible to show
that for loading in a discrete step the limiting value for the critical load is $\sqrt{2}-1$ and for quasi-static loading it is $3/8$. The remaining question,
therefore, is to find the rule of load transfer following a local failure that can achieve the global failure threshold in the closest
proximity to the above mentioned limits.
\begin{figure}[t]
\centering 
\includegraphics[width=9cm]{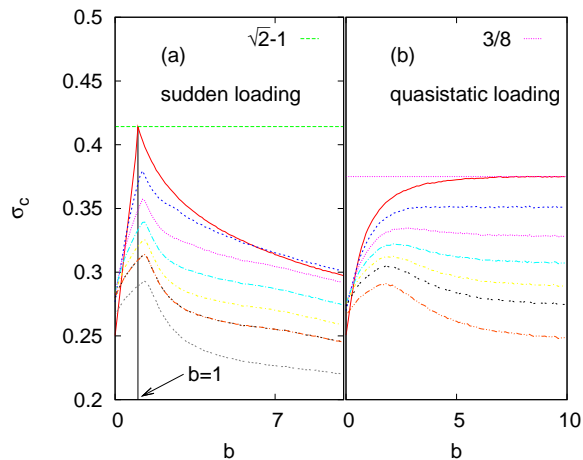}
   \caption{The phase diagram in the $b-\sigma_c$ plane ($b$ represents the anisotropy in the load redistribution process) is shown for (a) discrete step and (b) quasi-static loading for various fractional errors in the
knowledge of the threshold values of the individual fibers (curves from top to below are for $e=0.0, 0.1, 0.2, 0.3, 0.4, 0.5, 0.75$). The upper bounds for both cases are shown
which are reached for $b=1$ (a) and  $b \to \infty$ (b). From \cite{bs15}}
\label{maximize}
\end{figure}
Intuitively, it is clear that a higher share of load should be transferred to the fibers with higher capacity. Generally, it is
useful to assume that the transfer rule would be of the form $A(f_i-\sigma_i)^b$, where $f_i$ and $\sigma_i$ are respectively the
failure threshold and load of the $i$-th element, $A$ is an appropriate constant to ensure load conservation and $b$ is a parameter.

The dynamics, as discussed before, depends on whether the load is applied in a discrete step or gradually. The maximization of the
strength of the system would also, therefore, depend on the loading protocol. The only parameter to tune here is $b$. It is possible
to calculate analytically that the maximum strength is indeed achieved with this redistribution rule for $b=1$ for the discrete step
loading and $b\to \infty$ (practically achieved for $b\approx 10$) for quasi-static loading (see Fig. \ref{maximize}). 

An important information in implementing the redistribution rule is the exact knowledge of the failure thresholds of all the surviving elements.
This requirement may not be always fulfilled. Assuming that there is a (fractional) error $e$ in the knowledge of the failure thresholds, 
numerical simulations show (see fig. \ref{maximize}) that the redistribution rule still gives better results than a uniform redistribution. 
Therefore, in the situations where the load redistribution is controllable, the redistribution rule mentioned above give the best possible outcome.

We like to mention that cooperative dynamics appears in another class of 
Fiber Bundle Models  where fibers are treated as viscoelastic elements \cite{hkj02,j11,bk07}. The 
readers can go through \cite{crs20} (appearing in the same Research Topic: The Fiber Bundle) for 
a review on viscoelastic Fiber Bundle Models.

\section{Summary and conclusions}
One can easily see, the fiber bundle model (FBM), introduced
by  Peirce \cite{p26} in 1926 as a model to understand the
strength of composite materials, is an extremely elegant
one. As mentioned before, the model consists of a
macroscopically large number of parallel fibers/springs with linear elastic 
behavior 
and of identical length. The breaking thresholds, however, are different for 
each fiber and are drawn from a probability distribution. 
All these fibers/springs hang from a rigid horizontal platform. The load on 
the bundle is applied at the lower horizontal platform.
This lower platform has been assumed here to be
rigid, implying the stress or load-share per surviving
fibers/springs 
is equal, irrespective of how many fibers or springs might
have broken (equal load sharing or ELS scheme). It may be mentioned that we have not discussed
here the extensive studies on fiber bundle models with
Local Load Sharing (LLS) schemes, for which the readers
may be advised to consult Refs. \cite{RMP} and \cite{hhp15}
 and the `impregnated fiber bundle' models, for which the readers may be
referred to the Refs. \cite{new2,new3}.

As discussed in this review,  the failure dynamics of
the of FBM under ELS scheme of load sharing  have been
analyzed for long, both analytically as well as numerically
by several  distinguished groups of investigators from
engineering, physics and applied mathematicians. The
results may be briefly summarized as follows: After introducing the 
model, we have 
described the dynamics of the equal load sharing (ELS) fiber bundle model
in sec. II. Specifically, In this section we discuss and summarize works 
(Refs. \cite{hhp15,RMP,pc01,bpc03,ph07}) related to the cooperative failure 
dynamics in the ELS fiber bundle model having large number of fibers with 
different strength thresholds.  We start this section by describing the force 
displacement relation (load curve) when the bundle is stretched by an amount 
$x$. The maximum point of this curve gives the strength of the whole 
bundle. One can easily derive the strength of the bundle for different fiber 
threshold distributions. We have chosen uniform and Weibull distributions as 
examples and derive bundle’s strength as critical displacement ($x_c$)
and critical force ($F_c$). Next we describe how to formulate the dynamics of 
failure through a recursion relation in case of loading by discrete steps when 
fiber thresholds are uniformly distributed. The solution of the recursion 
relation at the fixed point gives some important information of the failure 
dynamics: Order parameter goes to zero following a power law as the applied 
stress values approaches critical value and both susceptibility and relaxation 
time diverge at the critical stress following well defined power laws. 
To check the universality of the failure dynamics we choose 
different type of fiber strength distribution (linearly increasing) and derive 
the fixed-point solutions. The exponent values 
of the power laws for order parameter, susceptibility and relaxation time 
variations are exactly the same as the model with a uniform distribution and 
therefore the failure dynamics in ELS fiber bundle model is universal. In 
addition, we present the exact 
solutions for pre and post-critical relaxation behavior which we believe is 
one of the most important theoretical developments in this field. 
At the last 
part of this section we present an analysis on the avalanche statistics for 
loading by fixed amount. Such a loading scheme  introduces a different 
mechanism for the avalanche sizes of 
simultaneous breaking of fibers. We discuss using analytical calculations 
that the exponent of the avalanche size distribution ($P(S)$) for discrete 
loading would be $-3$, 
which is different ($-5/2$) from the same  in case of quasi-static loading 
situation \cite{HH}. 

In section III, we summarize some recent developments
(Refs. \cite{Coleman,Ciliberto,Roux,pc03,pcc13,reiw_2009}) on the cooperative 
dynamics of
noise-induced failure in ELS fiber bundle models. In
addition to applied stress, the noise factor plays
a crucial role in triggering failure of individual
fibers. The trick here is how to define the failure
probability of individual fibers as a function of
applied/effective stress and the noise level. Normally
noise-level remains constant during the entire failure
process, but the stress level increases gradually due
to stress redistribution mechanism. The choice of the
probability function should satisfy the fact that
without noise factor the noise-induced failure model
must reproduce the classical failure scenario (discussed
in section II). We start this section by presenting
a noise-induced failure probability for individual
fiber failure. The choice of stress and noise level
dictates whether the system is in continuous breaking
regime or in intermittent breaking regime. Through a
mean-field argument, one can easily find out the phase
diagram separating these two regimes (Eq. (\ref{noise_ph_boundary}), 
Fig. \ref{fig:fg1}).
Apparently, the continuous breaking regime is easy to
analyze. For homogeneous bundle, where all the fibers
are identical (strengths are same), one can write down
the failure dynamics as a recursion relation (Eq. (\ref{noise_rec})).
The solution gives an exact estimate for the failure
time (steps) as a function of applied stress
($\sigma$) and noise level ($T$) (Eq. (\ref{noise_tf})). Simulation results
show perfect agreement with the theoretical estimates
(Fig. \ref{fig-homo}). When we consider a strength distribution
among the fibers in the model, it becomes extremely
difficult to construct the recursion relation for the
failure dynamics. One reason could be that during
the failure process the strength distribution gets
changed with time. However, the simulation results
(Fig. \ref{fig-hetero}) for the failure time of heterogeneous
bundles follow similar variation with applied stress
and noise level with an extra noise factor (Eq. (\ref{noise_tf_het})).
Next we discuss the other regime, i.e., the intermittent
failure regime where there are waiting time between two
failure phases. The distribution of waiting time is the
most important aspect in this regime. Simulation results
on homogeneous and heterogeneous bundles show that the
waiting time distribution follows a Gamma distribution
(Eq. (\ref{gamma_dist})) and a data-collapse confirms the universal
nature of such distribution function (Fig. \ref{fig:fg2}).
Surprisingly, waiting time distribution from earthquake
time series (California catalog) seem to follow similar
Gamma distribution (Fig. \ref{fig:fg3}).

In section  IV, we have considered self-organized fracture
front propagation in a fiber bundle model where the
fracture front adjusts its size in a self-organized way
to meet the increasing load on the bundle and several
features of the  self-organized dynamics can still be
analyzed in a mean field way; see, e.g., Fig. \ref{avalanche-comb} for the
avalanche size distribution which fits well with
$D(S) \sim S^{-3/2}$.

As mentioned already (in section II), the universality
class of the dynamics of fixed load increment during the
ongoing dynamics of failure in the bundle (until its
complete failure) will be different from that for the
quasi-static (or weakest link failure type) loading
during its dynamics. And, as discussed in section V,
it is given by the Flory statistics for linear polymers,
when fracture dynamics in the bundle is mapped to
turbulence and one utilizes the Kolmogorov type
dispersion energy cascades \cite{bc20}. In particular,
we obtained already (\cite{pbc02}; see Eqns. (\ref{eq:order-linear}) and (\ref{eq:kappa-linear})) the
order parameter exponent $\beta  = 1/2 = \gamma$, the
susceptibility exponent. Employing the Rushbrooke scaling
$2\beta + \gamma = d\nu$ (where $\nu$  denotes the
correlation length exponent), we get $d\nu  = 3/2$ here
in conformity with finite size scaling results. As
discussed in \cite{bc20} (see also the discussions in
section V), mapping the avalanche size distribution
(Eq. (\ref{avalanche_eq})) to the Kolmogorov energy dispersion
in turbulence (Eq. (\ref{kol_eq})), identifying $S$ with the
energy and inverse correlation length as the wave
vector $q$, we got the upper critical dimension $d_u$
for FBM in the ELS scheme to be $6$. This suggests
the correlation length exponent $\nu$ value here to
be $1/4$.

As discussed in this review, the absence of
stress concentrations or fluctuations around
the broken fibers allows mean-field type statistical
analysis in such Equal Load Sharing Fiber Bundle
Models. This feature of the models helped major
analytical studies for the  breaking dynamics and
also allowed precise comparisons with computer
simulation results.

\section{Acknowledgments}
This work was partly supported by the
Research Council of Norway through its Centers of Excellence funding
scheme, project number 262644. BKC is grateful to J. C. Bose Fellowship Grant 
for support.

\end{document}